\documentclass[prb,notitlepage]{revtex4-1}
\usepackage[a4paper,top=2.75cm,bottom=2.75cm,left=2.75cm,right=2.75cm]{geometry}
\usepackage{mathptmx}
\usepackage{bookmath}
\usepackage{graphicx}
\def\nicefrac#1#2{\frac{#1}{#2}}
\usepackage{hyperref}
\begin{document}
\title{\LARGE Majorana fermions as quanta of a superfluid vacuum}
\author{James A. Sauls}
\affiliation{Department of Physics and Astronomy, Northwestern University}
\date{December 9, 2013}
\maketitle
%--------------------------------------------------------------------------------------------------
\begin{center}
\begin{minipage}{0.8\textwidth}\large
%This news article on a thermodynamic signature of Majorana fermions confined on the surface of superfluid \Heb\ in the presence of a moving superfluid condensate was publsished in the Fall 2013 newsletter ``Dimensions'' of the Department of Physics and Astronomy, Northwestern University. 
%
This short article on a thermodynamic signature of Majorana fermions confined on the surface of a 3D topological superfluid in the presence of a moving superfluid condensate was circulated to a limited audience in Fall 2013.
Since recent experiments are looking at possible signatures of a gas of Majorana excitations confined in two dimensions on the surface of superfluid \Heb~\footnote{\large{\href{https://scholar.google.com/scholar?oi=bibs&cluster=6203946177679477885&btnI=1&hl=en}{Bound quasiparticle transport along the edges of superfluid $^3$He}}, D. Zmeev, S Autti, A Jennings, R Haley, G Pickett, Bulletin of the American Physical Society, 2022} it seems timely to post this earlier perspective to arXiv.
\end{minipage}
\end{center}
\bigskip
%--------------------------------------------------------------------------------------------------
\LARGE

In a perspective article ``Majorana returns'', published in Nature Physics,\footnote{\large\emph{Majorana Returns}, Nature Physics 5, 614 - 618 (2009), Frank Wilczek.} Franck Wilczek wrote ``In his short career, Ettore Majorana made several profound contributions. One of them, his concept of ‘Majorana fermions’ - particles that are their own antiparticle - is finding ever wider relevance in modern physics.''

Majorana's idea, published in 1937, was born from Paul Dirac's relativistic quantum theory of the electron, the centerpiece being the \emph{Dirac equation}, which not only accounted for electron spin of $\nicefrac{1}{2}$ (in units of Planck's quantum $\hbar$), but led to the prediction of the positron - the anti-particle of the electron with opposite electrical charge.
Majorana found a new equation describing neutral, spin $\nicefrac{1}{2}$ particles that are their own anti-particle, and he speculated that the neutrino might be such a particle. Years later the discovery that the various flavors of neutrinos have mass, and can transform into one another, combined with theories for the spectrum of particles at energies near the Higgs scale, have fueled new generations of experiments designed to search for definitive signatures that neutrinos `are' - or `are not' - Majorana particles, as well as hypothesized supersymmetric partners to
the neutral gauge bosons of the standard model, all of which are Majorana fermions.

However, neutrinos, LHC and dark matter searches are not the only physics frontiers where Majorana's elusive particles have found renewed interest. The ``ever wider relevance'' that Wilczek refers to in his article are developments and discoveries in condensed matter where Majorana fermions \emph{emerge} as quanta at ultra-low energies in a number of quantum liquid and solid phases of matter.
Indeed since 2008 there has been an explosion in theoretical predictions and discovery of new materials - \emph{insulators, superconductors and superfluids} - in which Majorana's idea has found relevance. 

Among the physics frontiers in the search for Majorana fermions are the ultra-low temperature phases of the light isotope of Helium (\He) - the B phase of superfluid \He\ to be specific.~\footnote{\large\emph{The Superfluid Phases of $^3$He}, D. Vollhardt and P. W{\"o}lfle, Taylor and Francis, 1990.} Not long after their discovery it was realized that the superfluid phases of \He\ provided a paradigm for one of the conceptual cornerstones of modern physics, namely the role of spontaneous symmetry breaking in quantum field theory and condensed matter physics.\footnote{\large\emph{The Universe in a Helium Droplet}, Oxford University Press, 2009, Grigory E. Volovik.}
The developments in this field led to separate Nobel prizes for the discovery of the phases by Douglas Osheroff, Robert Richardson and David Lee in 1996, and for theoretical developments leading to the identification of these macroscopic quantum states of matter by Anthony J. Leggett in 2003.

\begin{figure}[t]
\includegraphics[width=0.75\linewidth,keepaspectratio]{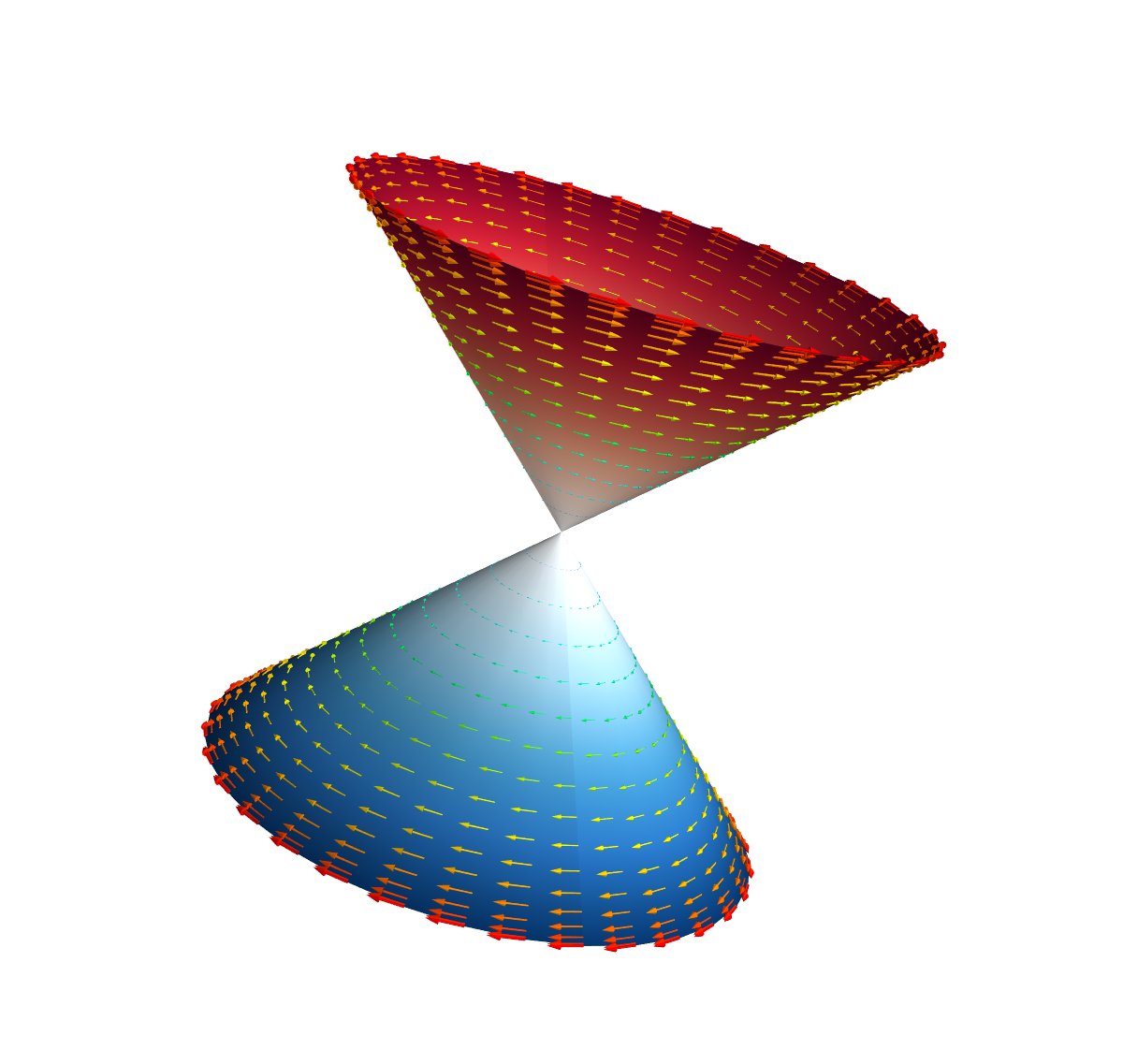}
\caption{\large
Helical spin structure of the Majorana spectrum of superfluid \Heb. The spin a Majorana excitation 
is locked along the direction of momentum $\vp$. At $T=0$ the occupied negative energy states give rise 
to a ground-state spin current, but zero mass current, confined on the vacuum-superfluid surface.
}
\label{fig-Majorana_Cone}
\end{figure}

But only since 2008 has it been widely appreciated that the ground state of this quantum liquid has remarkable properties connected with the mathematics of topology that imply that Majorana fermions exist as quanta. What many physicists believe they understand about superfluid \He\ suggests that Majorana fermions emerge from the ground state of superfluid \He\ - what is aptly dubbed the ``superfluid vacuum'' - as low-energy quanta, confined on a surface or a boundary of superfluid \He.~\footnote{\large\emph{Multifaceted properties of Andreev bound states: interplay of symmetry and topology}, T. Mizushima and K. Machida, Phil. Trans. R. Soc. A. {\bf 376}, 20150355 (2018).} 
So how might these exotic particles be detected?
Several theoretical proposals focus on the spin-momentum locking of the Majorana fermions, or the \emph{spin current} carried by Majorana particles, illustrated in Fig.~\ref{fig-Majorana_Cone}. However, detecting spin currents turns out to be complicated, sufficiently so that the elusive Majorana particles have so far remained just that.

%--------------------------------------------------------------------------------------------------
%--------------------------------------------------------------------------------------------------

A paper published in Physical Review suggests another route to detection.\footnote{\large\emph{Majorana excitations, spin and mass currents on the surface of topological superfluid \Heb}, Phys. Rev. B 88, 184506 (2013), Hao Wu and J. A. Sauls.}
The idea is based on the energy-momentum relation for the Majorana particles, $E = c|\vec{p}|$, where the ``light speed'' of the superfluid vacuum is $c\approx 5\,\mbox{cm/sec}$.
The superfluid vacuum offers a method for detecting these quanta at ultra-low temperatures by utilizing the Doppler shift of the Majorana fermions if the superfluid vacuum is set in motion. 

\begin{figure}[t]
\includegraphics[width=0.75\linewidth,keepaspectratio]{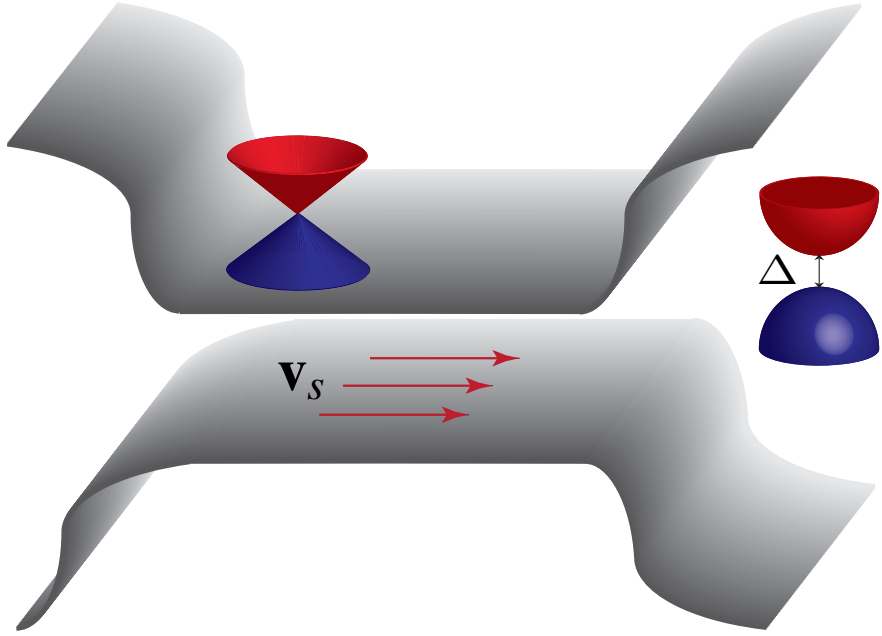}
\caption{\large Majorana fermions confined on the boundary of \He\ have a massless dispersion,
$E=c|\vp|$, shown as the Majorana `cone', while Dirac fermions with mass $\Delta$, are the quanta in
bulk \He. Relative motion of the superfluid vacuum ($\vec{v}_s$) leads to a Doppler shift in the
energy of Majorana fermions. }
\label{fig-superflow-channel}
\end{figure}

The fundamental property of a superfluid is that it supports a `persistent current', i.e. mass flow without dissipation - the same phenomena responsible for superconductivity, i.e. the flow of electricity without Joule losses.
For \He\ confined within a channel (Fig. \ref{fig-superflow-channel}), motion of the superfluid relative to the channel walls leads to a Doppler shift in the spectrum of Majorana particles confined on the walls, $E \rightarrow E' = c|\vec{p}| - \vec{p}\cdot\vec{v}_s$, where $\vec{v}_s$ is the velocity of the superfluid vacuum relative to the channel walls.

%--------------------------------------------------------------------------------------------------
%--------------------------------------------------------------------------------------------------

The fraction of mass that can participate in the persistent current depends on the energy and momentum distribution of fermi particles present - in this case the Doppler shifted Majorana fermions - and the absolute temperature, $T$, which controls the number of thermally excited Majorana fermions.
The Doppler effect leads to asymmetry in the thermal populations of Majorana excitations that are co-moving and counter-moving relative to the superfluid vacuum, and thus to a reduction in the persistent current.
The linear dispersion of the Majorana excitations, their fermi statistics, combined with their confinement in two dimensions leads to a unique signature: a power law dependence for the reduction in the mass fraction of the persistent current with temperature, $\delta n_s/n \propto - (T/c\,p_f)^3$, where $p_f$ is the fermi momentum, $c$ is the speed of the Majorana particles and $c\,p_f = \Delta$ is the mass gap for fermions in the bulk of the superfluid vacuum.
Just as the radiation flux from matter in thermal equilibrium, proportional to $T^4$, is a signature of the thermal distribution of photons, the quanta of the electromagnetic vacuum,
Majorana fermions reveal their presence as a mass flux, proportional to $T^3$, in response to the motion of the superfluid vacuum.

While we await the discovery of Majorana fermmions, new experiments are being planned and theorists are exploring the ``open range'' of ramifications of matter in which Majorana particles naturally emerge.
Whether or not all of the ideas spawned by Majorana's return to center stage will lead to deeper
insight into quantum correlations, or to a new generation of quantum computers is far from being
answered, but one thing is known.
If Majorana fermions are \emph{not} found in superfluid \He\ then of one of the most successful theories of twentieth century physics - the theory of quantum liquid phases of \He\ will need a pretty serious adjustment.
I'm betting that Majorana's particles are zipping about the surface of \He, and that there is a remarkable two-dimensional quantum liquid waiting to reveal its secrets.~\footnote{\large\underline{Update:} The detection of Majorana fermions in \Heb\ was reported based on the scattering of thermally excited Majorana excitations off electrons moving at different depths below the surface of superfluid \Heb. For a detailed discussion of these experiments and the supporting theory see the review: \href{https://doi.org/10.1007/s10909-022-02677-0}{\it Observation of Majorana Bound States at a Free Surface of $^3$He-B}, Hiroki Ikegami and Kimitoshi Kono, J. Low Temp. Phys. 195, 343–357 (2019).}
And who knows, maybe we will find out that the larger universe we are embedded in
is just as extraordinary as the superfluid vacuum!
%---------------------------------------------------------------------------------------------------

\end{document}